\documentstyle[11pt,aaspp4]{article}

\def\kms{km~s$^{-1}$}
\def\simlt{\lower.5ex\hbox{$\; \buildrel < \over \sim \;$}}
\def\simgt{\lower.5ex\hbox{$\; \buildrel > \over \sim \;$}}
\def\hb{\hfill\break}

\def\msol{{$M_\odot$}}

\def\vlsr{v_{\rm LSR}}

\def\mhtwo{M({\rm H_2})}

\def\nhtwod{n({\rm H_2})}

\def\vs2{v_{s2}}

\def\ratioto{^{12}R_{2-1/1-0}}
\def\ratiott{^{12/13}R_{2-1}}
\def\ratioot{^{12/13}R_{1-0}}

\def\xdvdr{X({\rm ^{12}CO})/(dv/dr)}

\def\lcoone{L_{10}(^{12}{\rm CO})}
\def\lcotwo{L_{21}(^{12}{\rm CO})}
\def\xco{X(^{12}{\rm CO})}

\lefthead{Koo et al.}
\righthead{SHOCKED MOLECULAR GAS IN THE SUPERNOVA REMNANT HB~21}

\begin{document}

\title{ SHOCKED MOLECULAR GAS IN THE SUPERNOVA REMNANT HB 21}

\author{Bon-Chul Koo}
\affil{Astronomy Program, SEES, Seoul National University,
Seoul 151-742, Korea;\\
koo@astrohi.snu.ac.kr}
\author{Jeonghee Rho \& William T. Reach}
\affil{Infrared Processing and Analysis Center, California Institute of 
Technology, Pasadena, CA 91125}
\author{JaeHoon Jung}
\affil{Korea Astronomy Observatory, Daeduk, Korea} 
\author{Jeffrey, G. Mangum}
\affil{NRAO, Tucson, AZ 85721-0665}

\begin{abstract}

We have carried out $^{12}$CO J=2--1 
line observations of the supernova remnant (SNR) HB~21 in order to search for 
evidence of interaction with molecular clouds. 
We mapped the eastern half ($80'\times 110'$) of the SNR almost completely.
Molecular gas appears to be distributed mainly 
along the boundary of the SNR, 
but the overall distribution has little correlation either with 
the distortion of 
the SNR boundary or with the distribution of radio brightness. 
Along the eastern boundary, 
where the SNR was considered to be interacting with molecular clouds 
in previous studies, we have not found any strong 
evidence for the interaction. 
Instead we detected broad 
(20--40~\kms) CO emission lines in the northern and 
southern parts of the SNR. In the northern area, the broad-line emitting cloud 
is composed of a small ($\sim 2'$ or 0.5~pc), very bright,
U-shaped part and several clumps scattered around it. There is 
a significant enhancement of radio emission with flat ($-0.28\pm 0.17$) 
spectral index possibly associated with 
this cloud. In the southern area, the broad-line emitting cloud is 
filamentary and appears to form an elongated loop of $\sim 30'$ in extent.
Small ($\simlt 1'.2$ or 0.3~pc), bright clumps 
are seen along the filamentary structure. 
We have obtained sensitive J=1--0 and J=2--1 spectra of
$^{12}$CO and $^{13}$CO molecules toward several peak positions. 
The intensity of $^{12}$CO J=2--1 emission is low ($T_{mb}<7$~K) and the 
ratio of $^{12}$CO J=2--1 to J=1--0 integrated intensities 
is high (1.6--2.3), which suggests that the emission is 
from warm, dense, and clumpy gas. 
We have applied an LVG analysis to derive their physical parameters.
The detected broad CO lines are believed to be emitted from 
the fast-moving molecular gas swept-up by the SNR shock. 
The small ($\simlt 20$~\kms) shock velocity suggests that the shock 
is a non-dissociating C-shock. We discuss the correlation of 
the shocked molecular gas with the previously detected, 
shocked atomic gas and the associated 
infrared emission.

\end{abstract}


\section{INTRODUCTION}

The number of supernova remnants (SNRs)
with convincing evidence for interaction with ambient molecular clouds 
has increased considerably in recent years. The evidence 
ranges from a simple morphological relation to the detection of 
broad and/or shock-excited emission lines from various molecules.
Although circumstantial evidence could be very suggestive, it is the molecular 
lines from the shocked gas that are essential for understanding
the physical and chemical processes associated with the molecular shock. 
In this regard, there are still only a few SNRs adequate for the study of 
molecular cloud-shock interaction; perhaps 
W28 (Arikawa et al. 1999), W44 (Seta et al. 1998), W51C (Koo \& Moon 1997), 
3C391 (Reach \& Rho 1996, 1999), and the classical source IC~443 
(DeNoyer 1979; Tauber et al. 1994 and references therein). 

In this paper, we report the discovery of broad emission lines 
from the shocked CO gas in HB~21.
HB 21 (G89.0+4.7) is one of those SNRs
with mixed morphology, e.g., shell-like in radio and center-filled in
the X-ray (Rho \& Petre 1998), where the center-filled, thermal X-ray 
emission is suggested to be due to interaction with molecular clouds.
It has a nearly complete radio-continuum shell with an angular extent
of $\sim 120' \times 90'$ (Hill 1974; Tatematsu et al. 1990, hereafter T90).
The shell is elongated along the northwest-southeast direction. The
brightness distribution of the shell is not uniform, but enhanced in 
scattered areas. Particularly noticeable features are 
the V-shaped northern boundary, a $\sim 30'$-sized loop structure in the south
central area, and  the one in the central region of the eastern boundary 
(see Fig. 1). 
Optical nebulosity associated with HB~21 
has not been detected in H$\alpha$ or [SII] plates (van den Bergh 1978).
X-ray emission from HB~21 was
detected by Leahy (1987) and studied in detail by Leahy \& Aschenbach (1996).
The distance to the SNR is uncertain. We adopt 0.8~kpc 
following T90, which is the distance to the Cyg OB~7 complex 
(Humphreys 1978).

HB 21 has been a suspect for interaction with molecular clouds based on
its radio appearance and the ambient molecular clouds (Erkes \& Dickel 1969; 
Huang \& Thaddeus 1986; T90). 
Erkes \& Dickel (1969) suggested that the distorted boundaries with 
enhanced radio brightness might be the places where the SNR is interacting 
with dense ambient gas.
Huang \& Thaddeus (1986)
found that the giant molecular cloud associated with Cyg OB~7 appears to be
partially surrounding HB~21 (see also Dobashi et al. 1994). T90 
obtained a higher-resolution ($2.'7$) CO map of the eastern part of the
SNR and found that
the eastern boundary of the SNR appeared to be in contact with  
molecular and atomic clouds.
They also made a coarsely-sampled map of the regions with
enhanced radio emission and
detected molecular clumps. But no direct evidence for the
shocked {\it molecular} gas has been
detected for HB~21.  The search for OH masers, 
which are known to be an indicator
for the interaction between a SNR and molecular cloud, gave negative 
results too (Frail et al. 1996). 
 On the other hand, Koo \& Heiles (1991) detected shocked
H~I gas moving at 40 to 120~\kms\ in the southern part of the SNR, although 
the limited 
angular resolution ($36'$) hindered any detailed study of the shocked gas. 

We detected broad CO emission lines in the northern and southern parts of HB~21, but 
not in the eastern part. We have found no evidence suggesting that the molecular 
clouds in the eastern part are interacting with the SNR, although they appear to 
be located along the boundary of the SNR on the sky.
We summarize the observations 
and the results of molecular line observations in \S~2 and \S~3, respectively.
In \S~4. we discuss the implications of our results on the interaction between 
the SNR and molecular clouds. \S~5 summarizes the main results of our paper.

\section{OBSERVATIONS}

$^{12}$CO J=2--1 line observations were
carried out using the 12 m telescope of the
National Radio Astronomy Observatory \footnote{The National Radio
Astronomy Observatory is operated by Associated Universities, Inc., under
contract with the National Science Foundation.} at Kitt Peak
in 1999 June and 2000 January.
The FWHM of the telescope at 230~GHz was 27$''$.
We mapped the eastern half ($80'\times 110'$) of the SNR almost completely
using the OTF (On-The-Fly) observing technique. 
We used two 256 channel 
filter banks; one with 500 kHz, and the other with 
1~MHz resolution. We split each filter bank into two sections and observed two 
linear polarizations simultaneously. 
The velocity resolution and coverage of the 500~kHz filter bank were  
0.65~\kms\ and 83~\kms, while those of the 1~MHz filter bank were two times greater.
Typical system temperatures were 350--450~K. During the observing run in 
January 2000, we had some 
trouble because of telluric CO 
J=2--1 emission, which appeared at $\vlsr=4$--5~\kms. The emission was not cancelled 
out completely by usual position-switching observation 
and produced a hill-and-valley feature in spectra, the strength of which 
depends on elevation. We were able to avoid the contamination from telluric CO emission 
by averaging out the contaminated velocity channels because 
the telluric CO emission line is narrow (0.8~\kms) and the absolute strength 
of the `hill' and `valley' features are equal.

We also obtained sensitive spectra of 
$^{13}$CO J=2--1, $^{12}$CO J=1--0 and J=2--1 
lines toward several peak positions. 
For CO J=1--0 observations, 
we used 1 and 2 MHz filter banks, so that they have the same velocity resolution 
and coverage with those of CO J=2--1 line observations. We also used 
the millimeter autocorrelator (MAC) for the $^{12}$CO  
observations, which provided high resolution (0.25~\kms\ after smoothing) 
spectra.
We have converted the observed
temperatures ($T_R^*$) to the main-beam brightness
temperatures ($T_{mb}$) using the corrected main-beam efficiency provided by the
NRAO.

Additional observations of $^{12}$CO and $^{13}$CO J=1--0 lines 
were performed using the Taeduk Radio  Astronomy 
Observatory (TRAO) 13.7m telescope (HPBW=49$''$ at 115 GHz) in 2000 January and
March. An SIS receiver equipped with a quasi-optical sideband filter was 
used along with a 250 kHz, 256-channel filter bank.
The main beam efficiency was 0.41 at 115 GHz 
(Roh \& Jung 1999), and the pointing accuracy was better than 10$''$.
Typical system temperatures were about 
750~K at 115~GHz and 450~K at 110~GHz.

\section{CO RESULTS}
\subsection{Overall Distribution and Clouds in Eastern Area}

Figure~1 shows the distribution of the integrated intensity 
of CO J=2--1 emission. The velocity range is between $\vlsr=+3.9$~\kms\ and 
$-17.5$~\kms, which covers most of the emission. 
The overlaied contour map shows the 1420~MHz brightness 
distribution of HB~21 (T90).
The overall distribution of CO gas in Figure~1 
is not very different from the low-resolution map of T90. 
But Figure~1 shows much detailed structure because of its high resolution, high 
sensitivity, and complete sampling.
Figure~1 immediately shows that 
molecular gas is distributed mainly along the boundary of the SNR. 
(We observed the central area in Figure~1, which had not been 
covered in our CO J=2--1 observations, in CO J=1--0 line emission, and have 
detected only several small [$\sim 1'$] clumps other than some faint 
extension associated with the clouds in southern and eastern parts of the remnant.)
But the overall distribution has little correlation either with the 
distortion of SNR boundaries or 
with the distribution of radio brightness (cf. \S~3.2).
For the purpose of discussion, we divide the remnant into three areas 
(Fig. 2):
(1) eastern area (RA$>$ 20$^{\rm h}$ 47$^{\rm m}$) 
where three relatively large ($\simgt 15'$) clouds and 
several filamentary clouds are present,
(2) northern area centered at 
(20$^{\rm h}$ 46$^{\rm m}$, 51$^\circ$ 00$'$), 
where a small ($\sim 2'$), very bright U-shaped cloud is noticeable, 
and 
(3) southern area centered at 
(20$^{\rm h}$ 44$^{\rm m}$, 49$^\circ$ 50$'$), where 
clumpy and filamentary clouds with complicated structures are 
present. 
We detected broad (20--40~\kms) emission lines from the clouds in
the northern and southern areas, which will be discussed in detail 
in the next sections. 
In the following, we summarize the results on the eastern area.

In the eastern area, there are three clouds centered at declinations 
$\delta\simeq 
50^\circ$ 49$'$, $50^\circ$ 15$'$, and $\simlt 49^\circ$ $50'$. 
We call these three clouds by clouds A, B, and C   
following T90 (see Fig. 2 for the location of these clouds).  
Figure~3 shows the channel maps of the eastern area. 
The velocity ranges of 
the channel maps were chosen to show the essential features clearly.
Cloud A appears at $\vlsr=+9$ to $-6$~\kms\ and 
is composed of two velocity components 
centered at +6~\kms\ and $-2$~\kms, respectively.
The former component (6~\kms), which is seen in Figure~3a, is 
extended and the emission peaks at the southern part ($\delta\simeq 50^\circ$ 
45$'$) of the cloud, while 
the latter component ($-2$~\kms), which is seen in Figures~3b and 3c, 
is spatially confined and comprises 
the northern part of the cloud.
Their maximum brightnesses are 
$T_{mb,{\rm max}}=11~K$ and 7~K, respectively.
Cloud B appears at $\vlsr=+1$ to $-9$~\kms\ and is seen in Figures~3c and 3d. 
The south central part of the cloud, e.g., the region between 
$\delta=50^\circ 10'$--17$'$, is bright and appears to be connected to 
cloud C. Cloud C has two components at very different velocities, 
e.g., one at +2 to $-11$~\kms, which is seen in Figures~3c and 3d, 
and the other at +17 to +10~\kms, 
which is not shown in Figure~3 but has a distribution 
similar to the other velocity component.
According to the result of T90, cloud C extends to 
$\delta\simeq 49^\circ 40'$.

An interesting feature in Figure~3 is the semi-circular loop that appears 
above cloud B in Figure 3b. 
The ratio of the minor, which is along the NS direction, 
to the major axis is 0.8.
If it is at 0.8~kpc, the linear size of the semimajor axis would be $R_s=3$~pc. 
The velocity increases systematically from both ends to the northern 
top of the loop, which is consistent with an expanding loop.
The top portion is redshifted with respect to the both ends by $\simeq 3$~\kms. 
If we assume that the ellipticity 
is due to projection, then the expansion velocity would be 
$v_{\rm exp}\simeq 5$~\kms, so that the dynamical age of the ring is 
probably shorter than $R_s/v_{\rm exp}\simeq 6\times 10^5$~yrs. 
This is much greater than the age ($3 \times 10^4$~yrs, Koo \& Heiles 1991, 
scaled to 0.8~kpc adopted in this paper) of HB~21 and, therefore, 
the loop might not be associated with HB~21. 
We suspect that 
the loop is originated from some energetic phenomena in cloud B.
A faint 'V-shaped' structure that connects cloud B and the ends of the loop 
in Figure~3b seems to indicate the association of the two. 

\subsection{Northern U-Shaped Cloud}

The cloud in the northern area 
is composed of a small ($\sim 2'$), very bright, 
U-shaped cloud and several clumps scattered around it 
(Fig. 1; see also Fig. 9 for enlarged view). 
Figure~4 shows its velocity structure.  
There are several points to be made from Figure~4: First 
the U-shaped cloud is composed of several clumps, whose 
central velocities shift systematically from +3 to $-6$~\kms\ 
as we move from NE to NW along the structure. 
The integrated intensity attains a maximum at 
(20$^{\rm h}$ 46$^{\rm m}$ 03$^{\rm s}$.2, 51$^\circ$ 00$'$ 00$''$), 
which we call HB21:BML-N1 (Broad Molecular Line -- Northern Position 1), or 
simply N1.
Second, 
there are several other clumps in the field. These clumps, except the one 
near the southeastern corner at $\vlsr=-13$~\kms, 
appear over a wide ($>10$~\kms) velocity range.
Among them, the one at 
(20$^{\rm h}$ 45$^{\rm m}$ 55.$^{\rm s}$0, 51$^\circ$ 03$'$ 30$''$), 
which we call HB21:BML-N2 (or N2), 
appears over the widest ($\simeq 30$~\kms) velocity interval. 
Third, there is a diffuse emission at $+2$~\kms\ to the NE of the cloud. 
Its line is narrow (2--3~\kms) and it is part of a large ($\sim 20'$) 
cloud that appears to be connected to cloud A. We consider that 
the clumps aligned along the NE-SW direction in Figure~4 are associated and 
call them cloud N, i.e., 
cloud N does not include the diffuse emission 
in the northeastern area and the clump in the southeastern corner.
(cf. T90 detected in this area only the diffuse molecular gas at 
$\vlsr=1$ to 6~\kms\ and called it cloud D.)  

As can be expected from Figure~4, most clumps in cloud N have broad emission 
lines. As an example, we show the spectra of N1 and N2 
in the top frames in Figure~5, where we see that 
the spectrum of N1 is box-shaped and its full width (at zero intensity) is 30 km/s, 
while that of N2 
is asymmetric and extends from $-21$ to +11 km/s. For comparison, 
the spectrum of the diffuse, extended structure 
in the northeastern part of this 
area has narrow (2--3~\kms) emission lines centered at $+2$~\kms, 
a sample of which is shown in the right bottom frame in Figure~5. 

We have obtained sensitive J=1--0 and J=2--1 spectra of 
$^{12}$CO and $^{13}$CO molecules at the two peak positions, N1 and N2, and 
Figure~5 shows the spectra. The molecule and transition are marked in each 
spectrum. The second spectrum from the top is 
$^{12}$CO J=2--1 emission convolved to the J=1--0 beam size ($55''$) to be 
compared with the J=1--0 spectra. The difference between the top and convolved spectra 
indicates that some velocity components, e.g., 
the narrow component 
centered at +4~\kms\ of N1 and the broad component at $-8$~\kms\ of N2,  
are confined to small areas. 
By comparing the J=1--0 and the convolved J=2--1 spectra,
we notice that the ratio of J=2--1 to J=1--0 intensities is high and that 
it varies over the profile:
For N1, the ratio is between 1.2 and 2.0 in the 
central parts of the spectrum, while it 
increases at the wings, e.g., $\sim 5$ at +5~\kms\ and $-$13~\kms. 
For N2, the ratio varies between 0.8 and 2.8, and it is higher 
between $-11$~\kms\ and 0~\kms. The ratios of the 
$^{12}$CO J=2--1 and J=1--0 integrated intensities 
$\ratioto$ are 1.6 and 1.7 for N1 and N2, respectively.
The $^{13}$CO J=2--1 line is clearly detected toward N1, while it is
marginally detected toward N2. For N1, the line has double peaks centered
at +1 and $-9$ \kms, while the $^{12}$CO J=2--1 line profile toward N1
is composed of several narrow peaks. The narrow peaks
might indicate that the emission is from several, unresolved subclumps.
Presumably, the $^{13}$CO emission might be from these
subclumps too, which is not apparent in the profile in
Figure~5 because of low signal-to-noise ratio and low velocity resolution.
(Note that the velocity resolutions of the $^{12}$CO J=2-1 and $^{13}$CO J=2-1
lines are 0.25 and 0.68~\kms, respectively.)
The ratios of $^{12}$CO J=2--1 to $^{13}$CO J=2--1 integrated line
intensities $\ratiott$ are $20\pm 3$ and $40\pm 14$ for N1 and N2,
respectively.
(The errors are statistical errors.)
Table ~1 summarizes the line parameters of the peak positions, i.e., 
their coordinate, 
velocity range ($v_{\rm min}$, $v_{\rm max}$), 
CO J=2--1 peak brightness temperature $T_{mb,{\rm max}}$, 
$\ratioto$, $\ratiott$, and the ratio of $^{12}$CO J=1--0 to $^{13}$CO J=1--0 
integrated line intensities $\ratioot$. 
The detailed line diagnostics based on the observed line parameters in 
Table~1 is discussed in \S~3.4. 

\subsection {Southern Filamentary Cloud}

In the southern part of the SNR, the emission is detected at $\vlsr=-35$ to 
$+20$~\kms. The velocity structure is shown in Figure~6.
At positive velocities, we see several clouds with narrow lines come and go, e.g., 
a diffuse cloud that extends $\sim 10'$ 
along the NS direction centered at (20$^{\rm h}$ 44$^{\rm m}$.5, 50$^\circ$ 00$'$)
between $\vlsr=0$ and +5~\kms. 
The CO distribution at negative velocities is fairly complicated: the 
distribution is filamentary, and small 
($\simlt 1'.2$ or 0.3~pc), bright clumps 
are seen along the filamentary structure. 
The filamentary structure, which we call cloud S, appears to form a 
loop of $\sim 30' \times 10'$ in extent, elongated along the NS direction. 
The eastern part of the loop is particularly clumpy and has a semicircular 
shape (see the channel map centered at $-10.1$~\kms). 

The clumps generally have broad ($\simgt 10$~\kms) lines. Among them, 
three clumps 
marked by crosses in Figure~6
have broadest (30--40~\kms) lines and we show the $^{12}$CO J=2--1 and 
J=1--0 spectra at their peak positions, 
which we call S1, S2, and S3 from east to west (see Fig. 2), in Figure~7. 
Again we show the convolved J=2--1 spectra 
together, although the line shapes do not change 
significantly by convolution toward these peak positions. 
We have obtained some sensitive J=1--0 and J=2--1 spectra of 
$^{12}$CO and $^{13}$CO molecules at these peak positions, and 
the line parameters are listed in Table~1. Note that $\ratioto$(=1.7--2.3) and 
$\ratiott(=28\pm 4$ toward S1) 
are similar to those of northern positions, while $\ratioot$'s, although 
they have large uncertainties, appear 
to be much greater than that of N1.

\subsection{Excitation Parameters of Broad Emission Lines}

The broad CO lines are presumably emitted from the shocked gas, where physical 
parameters vary greatly over a short distance scale. But 
still it would be worthwhile to estimate their excitation parameters
based on elementary considerations. First, the observed   
$\ratiott=20$--40 are significantly less than either the average ratio
($67.3\pm1.5$; Langer 1997) of $^{12}$C/$^{13}$C 
in the solar neighborhood or the terrestrial value (89), implying that  
the $^{12}$CO J=2--1 lines are {\it not} optically thin. 
If we adopt $^{12}$C/$^{13}$C=67 and assume that the emission is thermalized, 
then the optical depth for the $^{12}$CO J=2--1 line 
$\tau_{21}=1.1$--3.4. On the other hand, the $\ratioot$($\simeq 100\pm 50$) 
close to or greater than 
the terrestrial value imply that $^{12}$CO J=1--0 lines are optically thin, except 
at N1 where J=1--0 and J=2--1 lines appear to have comparable optical depths.
This, however, is not conclusive because of large uncertainties associated with 
$\ratioot$.
Second, the large values of $\ratioto=1.6$--2.3 
imply that the broad-line emitting region is warm and dense. 
For typical molecular clouds, where the J=2 level is subthermally excited, 
the ratio is usually less than unity. 
For example, molecular clouds in the local arm exhibit ratios ranging from
0.53 (Taurus) to 0.75 (Orion A) (Sakamoto et al. 1994, 1997).
Our spectra of ambient gas also show this, e.g., see the spectra toward S2 in Fig. 7 where 
the narrow component at +2~\kms\ has $\ratioto=0.72$.
But the large ratio is common for the shocked molecular gas in SNRs (see \S~4.1). 
Third, the low ($<7$~K) brightness temperature of J=2--1 lines, 
regardless of their 
moderate optical depths, imply that the emitting region must be clumpy, i.e., 
composed of subclumps, and 
the emission is beam-diluted. 
From the above considerations, we may conclude that the broad emission lines are 
from warm, dense clumps with significant column densities so that 
the 2--1 lines are optically thick.

We have applied the large-velocity-gradient (LVG) model 
(Scoville \& Solomon 1974; Goldreich \& Kwan 1974) to our broad CO lines 
in order to derive their excitation parameters. 
The model assumes an uniform, spherical
cloud with a constant velocity gradient ($v(r) \propto r$).
If the lines are emitted from the 
shocked region where temperature and density vary greatly, 
the resulting parameters may be considered as `average' values.
Since we have found that the emission is beam-diluted, we have used 
the line ratios, $\ratioto$ and $\ratiott$, instead of 
brightness temperatures to determine the excitation parameters.
According to our LVG analysis,
the observed ratios are possible for $T_k\ge 50$~K. 
Figure 8a shows the result of our model computations when 
$T_k=100$~K, where curves of constant $\ratioto$ and constant $\ratiott$ are drawn 
in ($\xdvdr$, $\nhtwod$) plane. $\xdvdr$ is the fractional abundance of 
$^{12}$CO relative to H$_2$ ($\xco$) per unit velocity gradient interval. 
The asterisks (*) mark the observed ratios toward the peak positions where both 
ratios are obtained, i.e., N1, N2, and S1.
According to Figure~8a, 
$\nhtwod=(3$--$7) \times 10^3$~cm$^{-3}$ and 
$\xdvdr\simeq (1$--$4) \times 10^{-6}$
 pc~(\kms)$^{-1}$\ at the three peak positions. 
There are multiple choices for N1, e.g., the same ratios are obtainable when 
$\nhtwod=3\times 10^5$~cm$^{-3}$ and $\xdvdr=4 \times 10^{-7}$ pc~(\kms)$^{-1}$. 
We adopt the 
lower density because it is comparable to the densities in the other peak positions and 
because the density of $>10^5$~cm$^{-3}$ appears to be
 too high for the CO emission to explore.
If the temperature becomes higher, both $\nhtwod$ and $\xdvdr$ need to be greater.


In Figure 8b, we plot the {\it expected} CO J=2--1 
radiation temperature $J_\nu (T_b)\equiv (h\nu/k_B)/[\exp(h\nu/k_B T_b)-1]$ 
which is just the brightness 
temperature when $h\nu \ll k_B T_b$ and the expected 
$\ratioot$ from the same LVG model. Note that the expected radiation temperatures are
much greater than the observed main-beam brightness temperature. 
We have estimated beam filling factors of 
(7.7--8.8)$\times 10^{-2}$ 
from the ratio of these two brightnesses.
We have estimated 
the CO column densities $\simeq (2.4$--$11) \times 10^{17}$~cm$^{-2}$ 
at these peak positions by  
$\nhtwod [\xdvdr] \Delta v$, 
where $\Delta v=14$--18~\kms\ is the 
velocity width. 
The excitation parameters derived from the LVG analysis are listed in Table~2.
Note that the $\ratioot$ expected from the LVG model 
differ from the observed ones: 
At N1, the observed value is small by a factor of 2, while, at S1, it is 
large by a factor of $2\pm 1$. Considering the weakness of 
$^{13}$CO J=1--0 lines and various uncertainties associated with 
different telescopes, 
however, it is not obvious if this difference is critical. 

We have made a crude estimate of the mass of the broad-line clouds 
as follows. 
If the CO J=1--0 line emission is optically thin, then $\mhtwo$ can be obtained from
the CO J=1--0 luminosity $\lcoone$ by 
$\mhtwo=\lcoone m_{\rm H_2} /[h \nu_{10} A_{10} f_{J=1} \xco]$ where 
$f_{J=1}$ is the fraction of CO molecules at $J=1$ level and the other 
coefficients have their usual meanings. In our case, CO J=1--0 emission has less optical 
depth than the J=2--1 emission, but is 
not very optically thin, so that the above formula might yield an underestimate.
What we have is the luminosity of 
CO J=2--1 emission $\lcotwo$, which has moderate optical depth. 
But, since $\ratioto=1.6$--2.3 at the peak positions, we may obtain 
$\lcoone$ by assuming that $\lcoone=(1/2)(\nu_{10}/\nu_{21})^3\lcotwo$ where 
$\nu_{10}$ and $\nu_{21}$ are CO $J=$1--0 and $J=$2--1 line frequencies respectively. 
Finally we assume $f_{J=1}=0.2$, which is a mean value of those (0.15--0.26) 
at the three peak positions obtained from the LVG analysis.
We have found that the H$_2$ masses of  
clouds N and S are $\sim 8$~\msol\ and $\sim 55$~\msol, respectively.
The mass of the central U-shaped part of cloud N is $\sim 3$~\msol\ while the masses of the 
small clumps in cloud S are $\sim$1~\msol.

\section {INTERACTION BETWEEN HB 21 AND MOLECULAR CLOUDS}

\subsection {Evidence for the Interaction}

Broad CO emission lines with large $\ratioto$  
in clouds N and S strongly suggest that they are 
being shocked. The observed velocity width is as large as $\sim 40$~\kms.
Note that, 
toward this direction ($\ell=89^\circ$), the LSR velocity permitted 
by the Galactic rotation is $\simlt 0$~\kms, so that 
it is not impossible for broad lines to be produced by molecular clouds 
accidentaly aligned along the line of sight. 
But it is highly improbable that such alignment (over a few kpcs)
occurs in very small (1--2$'$) areas on the sky. 
We also searched for protostellar candidates around the 
broad line emitting regions using the 
{\it Infrared Astronomical Satellite} ({\it IRAS}) Point Source Catalog, 
because broad lines can be emitted from  
the high-velocity gas associated with protostellar object too.  
We did not find any suspicious sources except one, IRAS 20444+4954, which is
located close to the S2 clump, i.e., at ($29''\pm14''$, $68''\pm11''$) from
the peak position in Table~1. The source has been detected in two IRAS wavebands,
i.e., 60 and 100~$\mu {\rm m}$, with flux densities of $F_{60\mu {\rm m}}=0.99$~~Jy and
$F_{100\mu {\rm m}}=12.97$~Jy. We have found that the source is located
within a small ($\sim 1'\times 2'$), bright ($T_{mb}\simeq 6$~K) CO J=2--1 core
at $\vlsr\simeq 1$~\kms, so that it might be a young stellar object associated
with the $\sim 10'$-sized, diffuse cloud in the northern part of the +2.9~\kms\ map in
Figure~6, not with the S2 clump. Also the velocity of the S2 clump is similar to
those of the other fast-moving clumps in this area, which suggests that they
have the common origin. Therefore, the broad lines that we detected are almost
certainly from the fast-moving molecular gas swept-up by the SNR shock in
HB~21. 


Another indication that the broad CO lines are from the shocked gas 
is their high $\ratioto=1.6$--2.3. As we have shown in \S~3.4, the high 
ratio implies that the emitting gas is warm and dense, which 
might be manifestation of shock.
Indeed high $\ratioto$ is 
a common property of the broad lines from the shocked molecular gas in SNRs:
All six SNRs that are known to have broad molecular 
emission lines, i.e., W28 (Arikawa et al. 1999), 3C391 (Reach \& Rho 1999), 
W44 (Seta et al. 1998), W51C (Koo \& Moon 1997), 
HB~21 (this paper), and IC~443 (e.g., van Dishoeck et al. 1993), 
have ratios greater than 1, which 
implies that the broad CO lines in these SNRs are all emitted from 
warm and dense, shocked gas. 
Meanwhile, the maximum brightness temperature ($< 7$~K) of CO J=2--1  
line in HB~21 is significantly less than those of other 
SNRs even if it was obtained with a higher spatial resolution (0.1~pc), 
e.g., it is 33~K for W28 (Reach \& Rho 2000b) 
and IC~443 (van Dishoeck et al. 1993) when observed with a resolution of 0.2~pc. 
The much smaller CO J=2--1 brightness temperature 
with comparable $\ratioto$ imply that either the shocked gas in HB~21 
is composed of much smaller clumps or is less dense.

On a large scale, HB~21 appears to be in contact with a giant molecular cloud (GMC) 
along its eastern boundary 
(Huang \& Thaddeus 1986; Tatematsu et al. 1990). 
The GMC is 130~pc$\times$70~pc in extent  
(Dobashi et al. 1994), and the structure that we call 
clouds A, B, and C defines 
the western boundary of the GMC.
T90 inspected the correlation between 
these clouds and the SNR in detail, and concluded that 
cloud A might be interacting with the SNR because it is located where 
the radio continuum boundary of the
SNR is distorted. On the other hand, they concluded that clouds B and C might 
not, because there is no indication of the interaction in the 
distribution of radio brightness. 
According to our high-resolution observations, however, 
there is little relationship between the 
boundaries of cloud A and the SNR.
Instead, since the velocity of the ambient molecular gas around HB~21 
might be negative (see \S~4.3), we consider that clouds B and C have 
a better chance of interaction.
But we have detected broad CO lines toward none of these eastern clouds.
Even if a SNR is interacting with a molecular cloud, 
the broad lines may be absent, however.  
3C~391 (G31.9+0.0), which appears to be located at the edge of a large molecular cloud, 
for example, has no broad CO lines along the interface 
(Wilner, Reynolds, \& Moffett 1998; Reach \& Rho 1999).
But strong [OI] 63~$\mu$m emission 
has been detected near the interface, which indicates that the SNR is interacting 
with the molecular cloud (Reach \& Rho 1996). This would happen 
if the shock is dissociative and molecules have not reformed.

\subsection {Enhanced Radio Emission Possibly 
Associated with Northern U-Shaped Cloud}

Figure~9 shows an enlarged view of the northern area overlaied with 
a 325~MHz radio continuum map of the SNR.
Note that there is an enhanced radio continuum emission 
elongated along the NE-SW direction in the central area.  
Its peak position falls exactly inside the U-shaped part of cloud N.
The positional coincidence 
suggests that the enhancement is possibly associated with cloud N.
We have derived a spectral index (flux$\propto \nu^{\alpha}$) of 
$-0.28 \pm 0.17$ for 
the radio emission associated with the U-shaped central part
using the 325~MHz and 1420~MHz maps.
The derived index has a large uncertainty because of the confusing 
``background" level, but 
it appears to be flatter than 
the mean spectral index $-0.4\pm0.03$ of HB~21 (Willis 1973). 
For cloud S, there is no obvious correlation between the CO emission and radio
continuum brightness, 
although the radio continuum appears to be bright around the cloud 
in general.

%
%

It is not obvious, observationally or theoretically, what determines 
whether or not 
radio synchrotron emission becomes enhanced when SN shock hits a dense cloud. 
Observationally, we see very limited correlation between radio continuum
brightness and shocked molecular gas in some SNRs, i.e., the 
shocked molecular gas is not usually associated with radio continuum enhancement 
and vice versa (see also Chevalier 1999).
In IC~443, for example, shocked molecular gas is distributed in a fragmentary,
flattened, ring, which partly overlaps with the radio continuum shell
(e.g., see Dickman et al. 1992 for the shocked molecular gas and Green 1986 for 
the radio continuum). But the radio continuum is brightest in the northeastern
part of the shell where there is no shocked molecular gas, and the radio continuum 
is not particularly bright toward the shocked molecular gas, perhaps except around 
the southern part of the molecular ring (see next).
In 3C~391, there is a shocked molecular clump in the southern part of the SNR,
but 
there is only a faint, local radio continuum peak at $\sim 1.5$~pc apart from the
shocked clump, whose association cannot be confirmed (Reach \& Rho 1999). In W28,
on the other hand, there is a ridge of radio continuum emission associated with
the shocked molecular gas (Arikawa et al. 1999). 
Theoretically, it could be either the dense cloud or 
the surrounding intercloud medium where synchrotron emission becomes enhanced.
If the shock propagating through the 
dense cloud is radiative, there will be a large 
compression of cosmic rays and magnetic field, which would 
increase the synchrotron emissivity 
(van der Lann 1962; Blandford \& Cowie 1982). But this mechanism may not work 
because 
the molecular shocks in old SNRs are not ionizing shocks and, therefore, 
high energy particles may escape 
from the shocked region (Draine \& McKee 1993; Chevalier 1999).
On the other hand, the shocked intercloud medium 
surrounding the cloud, particularly the medium behind the cloud,
could have enhanced 
synchrotron emissivity because of the increased magnetic field strength there 
(e.g., Jones \& Kang 1993; Mac Low et al. 1994).
In HB~21, the peak of the enhanced radio emission is 
located behind the shocked cloud (Fig. 8). 
It is also noteworthy that the cloud has a U-shape, which  
is similar to what we would expect when
a small cloud is swept up by a strong shock 
(e.g., Klein, McKee, \& Colella 1994; Mac Low et al. 1994). 
These morphological characteristics 
seem to suggest that the enhanced emission is not physically 
associated with 
the shocked cloud but with the shocked surrounding intercloud medium. 

In some respects, the region around cloud N in HB~21 is 
similar to the flat ($\alpha\sim-0.2$) 
spectral region in IC~443, which is also located behind a shocked molecular 
cloud (Green 1986; Keohane et al. 1997). In IC~443, 
Keohane et al. (1997) found that 
the flat spectral region is particularly bright in hard X-rays, which are 
most likely due to synchrotron radiation. They concluded that the enhanced 
hard X-ray emission and the flat spectral index is due to 
shock acceleration of cosmic rays behind dense clouds. 
More observational study is certainly needed for HB~21 in order to 
reveal the relationship 
between the northern U-shaped cloud and 
the enhanced radio emission.


\subsection {Nature of the Shock and Shock Parameters}

Cold molecular gas around HB~21 in general has central velocities 
between $\vlsr\simeq -8$~\kms\ and +6~\kms, which must be the velocity range of 
the preshock gas. In the areas where broad lines have been detected, there is 
diffuse molecular gas at positive (0 to +5~\kms) velocities 
(see Figures 4 and 6). If this gas represents the preshock gas, then, since the broad lines
are centered and spread out mostly at negative velocities, the gas should have 
been shocked and accelerated toward us {\it systematically}. 
If we take +3~\kms\ as the velocity of the preshock gas and if we take 
either the central or peak positions of the broad lines as the systematic velocities
of the shocked gas, then {\it the line-of-sight} velocities of the 
shock would be $\simlt 10$~\kms\ and $\simlt 20$~\kms\ for clouds N and S,
respectively. 
And, if we take $0.9R_s$ and $0.75R_s$ ($R_s$=the radius of the SNR) as 
their projected distances from the center of the SNR, then their deprojected 
velocities are $\simlt 23$~\kms\ and $\simlt 30$~\kms, respectively. 
Such coasting clumps are indeed theoretically expected for the old SNRs such as HB~21:
Molecular clumps swept up by a SNR blast wave are accelerated
by the shock propagating into the clumps 
and also by the ram pressure of interclump gas (e.g. McKee 1988).
The characteristic timescale for acceleration 
is $R/v_s$ where $R$ is the radius of the clump, 
which becomes
$\sim 1 \times 10^4$~yrs ($< t_{\rm age}\sim 3 \times 10^4$~yrs) 
for the clumps in HB~21 using
$R\simeq 0.25$~pc and $v_s\simeq $25~\kms. 
The wings of the broad 
lines may be attributed to the gas accelerated by shocks propagating 
from sides.
But one difficulty in this scenario 
would be that there is no clear 
correlation in the distributions of the broad lines 
between the preshock and postshock gases. 
In the southern area, for example, the preshock gas (2--3~\kms\ component) 
is distributed in a filamentary cloud extended 
along the NS direction, while the broad-line emitting clumps are distributed 
over a much wider area to the SW of this cloud. 
Also it appears rather awkward that there are no broad lines 
centered near the velocity of the preshock gas.
Alternatively, 
it could be the molecular 
gas at negative velocities that is associated with the SNR, and, 
in the northern and southern areas, most, if not all, of the ambient gas may have 
been shocked. In this picture, the shocked gas may be coasting too, but not necessarily 
at high speeds because the velocity of the preshock gas is somewhere within the 
velocity range of the broad lines, e.g., near the velocity of the peak.
The shock velocity determined from the line 
width is $\simlt 20$~\kms. Between the two possible interpretations, we prefer 
the latter because of the difficulty with the former mentioned above.
But, since the difference in the shock velocities 
between the two interpretations is $\simlt 10$~\kms,  
the following discussion remains valid basically even if the velocity of the preshock 
gas is 0 to +5~\kms.
(The derived properties of the shocked gas in \S~3.4 should remain valid too
in either case because the emission from the preshock gas might be in narrow lines and 
its contribution to the integrated intensity of broad lines is expected to be small.)  


The {\it observed} shock velocity is
$\simlt 20$~\kms. This is
less that the critical velocity
for the dissociation of molecules,
which is 25--50~\kms\ depending on preshock density and magnetic field strength
(Hollenbach \& McKee 1980; Draine, Roberge, \& Dalgarno 1983).
Hence, the shock might be a {\it non-dissociating C-shock}.
The observed integrated intensity of the CO J=2--1 emission is
(3--9)$\times 10^{-7}$~ergs cm$^{-2}$ s$^{-1}$ sr$^{-1}$ at the peak positions. 
If we consider the beam dilution (\S~3.4), the actual surface brightness 
may be greater by an order of magnitude, 
e.g., (4--10)$\times 10^{-6}$~ergs cm$^{-2}$ s$^{-1}$ sr$^{-1}$. 
This is much larger than the 
{\it angle-averaged} surface brightness predicted from 
shock model computations.
Draine \& Roberge (1984), for example,   
computed surface brightnesses expected for steady-state C-shocks propagating 
through molecular gas with different preshock conditions.
According to their result, the angle-averaged surface
brightness of CO J=2--1 emission varies from 
$1\times 10^{-7}$
to $2\times 10^{-6}$~ergs cm$^{-2}$ s$^{-1}$ sr$^{-1}$ 
for a 10--20~\kms\ shock propagating
through a molecular cloud with $\nhtwod=5\times 10^2$ to $5\times 10^3$~cm$^{-3}$. 
Larger shock velocity does not raise the surface brightness while higher preshock density 
may yield $\le 3\times 10^{-6}$~ergs cm$^{-2}$ s$^{-1}$ sr$^{-1}$.
The much higher surface brightness toward the peak positions would be possible if 
these are directions where we are observing the shock tangentially.


We want to briefly discuss the non-detection of OH
1720 MHz masers in HB~21, because such masers are known to indicate
interaction of SNRs with molecular clouds (Frail et al. 1996).
Firstly, the OH masers may require a very specific set of physical conditions
that might not be realized in HB~21. According to Lockett, Gauthier, \&
Elitzer (1999), the 1720 MHz masers arise only in C-shocks when $T=50$-125~K,
$n({\rm H_2})\sim 10^5~cm^{-3}$, and OH column density of
$10^{16}$--$10^{17}$~cm$^{-2}$. According to our result in \S~3.4, the density
of the shocked molecular gas in HB~21 appears to be much lower than
required. Secondly, it is not impossible that the OH maser emission, even if
present, had been missed in the survey by the Frail et al. (1996) who mapped
the SNRs in rectangular grids with full-beam (or $2 \times$ full-beam)
grid spacing. It would be worthwhile to search for OH masers toward the
shocked CO gas in HB~21.

Koo \& Heiles (1991) detected shocked H~I gas associated with HB~21. 
The shocked H~I gas moves at $\vlsr=42$--123~\kms\ and 
is confined to the southern part ($\delta=50^\circ$ 0$'$--50$^\circ$ 30$'$) 
of the SNR. The highest velocity component coincides with 
cloud S, although the angular 
resolution ($36'$) of the HI observation is too large for a detailed comparison. 
Koo \& Heiles (1991) assumed that the shocked H~I gas represents a cap portion 
of a large expanding H~I shell and derived a mean 
{\it ambient} H~I density of 3.7~cm$^{-3}$ 
(when scaled to 0.8~kpc adopted in this paper).
If the molecular shock has been driven by this H~I shell, then we 
can roughly estimate the density of the shell as follows:
We assume that the shocked molecular gas is confined to a 
thin slab and that 
the radiative H~I shell has an uniform density of $\rho_{rs}$.
Then it is 
straightforward to show that  $\rho_{rs}$ is related to the 
density of the molecular cloud $\rho_c$ by 
$\rho_{rs} \simeq \rho_c (v/v_{rs})^2$ (for $v_{rs}\gg v$) where
$v_{rs}$ and $v$ are the velocities of radiative shell and shocked molecular 
slab, respectively (e.g., see Chevalier 1999).
For HB~21, $v_{rs}\sim 130$~\kms\ and $v\sim 20$~\kms. And if we take 
the H$_2$ density of the cloud 
$n({\rm H_2})\sim 1\times 10^3$~cm$^{-3}$, the density of the 
H~I shell would be $n_{rs}({\rm H})\sim 47$~cm$^{-3}$. By comparing with 
the mean ambient density (3.7~cm$^{-3}$), this implies a compression factor 
$\beta\sim 13$ for the H~I shell. Such moderate compression would be 
obtained if the ambient magnetic field strength tangential to the shell 
is $B_0\simeq 2(8\pi\rho_0 v_{rs}^2)^{1/2}/(3\beta) \sim 
10$~$\mu$G where $\rho_0$ is the density of the ambient medium. 
(The equation is obtained by assuming that ambient magnetic 
field is uniform and magnetic pressure 
dominates the pressure in the shell. See Chevalier 1974 for a discussion.)
Alternatively, the fast-moving H~I gas 
could be the gas originally associated with the 
molecular clouds, i.e.,
the atomic and molecular shocks may be produced when the SNR shock hits a
large molecular cloud. In this case the
fast-moving H~I gas represents the swept-up
interclump medium or H~I envelope of molecular cloud.
High-resolution H~I observation is 
needed to reveal the relation between the shocked 
atomic and molecular gases.


\subsection{Infrared Emission from HB~21}

We used archival data from {\it IRAS} to search for infrared emission
associated with the remnant.
In his catalog of infrared emission from supernova remnants, Arendt (1989)
called HB~21 a "probable" infrared source at 12 and 60 $\mu$m, with total
fluxes of $180\pm 60$ and $800\pm 350$~Jy, respectively. The main source
of uncertainty is confusion with unrelated emission in the Galactic
plane, which cannot be easily separated in infrared images.
Using the {\it IRAS} {\it Sky Survey Atlas} ({\it ISSA}; Wheelock et al. 1993),
we created an image covering the region around HB~21 at 60 and 100 $\mu$m.
There is
extensive emission to the south, east, and west of the remnant, but with
no clear correlation with the radio or CO image. The region toward
the center of the remnant is relatively fainter than these edges, but the
northern part of the remnant is also faint. It is not possible to tell
whether the emission is related to a partial shell around the remnant
or just fluctuations in the background emission.

To search for infrared emission associated with the remnant in more detail,
we obtained a dedicated {\it IRAS} HIRES
(Aumann, Fowler, \& Melnyk 1990) image for a $2^\circ$ field centered on HB~21.
HB~21 is visible in the {\it IRAS} HIRES images at all four wavelengths
(12, 25, 60, and 100 $\mu$m).
The 60~$\mu$m image, where HB~21 is most prominent,
is shown in Figure~10.
Comparing the {\it IRAS} and CO images, it is evident that the
southern filamentary cloud, cloud S, is detected as a long arc,
with very similar location, shape and width.
The 60~$\mu$m surface brightness of
the filament is typically 7~MJy~sr$^{-1}$, and its structure is clumpy,
like that of the CO emission. But the peaks of CO and infrared emission do not
match in detail, suggesting that the 60 $\mu$m emission does not arise
from the exact same regions as the CO. In the northern area,
there is also a good correspondence between the
infrared and CO emission. The general correspondence in the
south and north, and partial overlap in the east, show that
many of the infrared features around the edge of HB~21 are related to
the remnant, although the infrared emitting regions differ in detail
from the CO emitting regions.

The nature of the infrared emission from HB~21 could be either
dust grains surviving the shock, or from spectral lines from shock-excited
gas, or both.
We have estimated the mean surface brightness and color of the CO clouds
by using several faint regions in the field as background. The results are
summarized in Table~3. The far-infrared color ratio, 
$I_{60}/I_{100}\simeq 0.20$, for clouds A, B, and C is almost identical to 
that of diffuse cirrus clouds in the solar neighborhood (Boulanger \& P\'erault 
1988).
This suggests that the infrared emission from clouds A, B, and C is most likely 
due
to dust heated at the surface of the molecular clouds by the interstellar 
radiation
field; specifically, it suggests that the infrared emission is not related to
shock fronts into the clouds.
On the other hand, clouds N and S have a significantly higher color ratio 
$I_{60}/I_{100}\simeq 0.27$.
This enhanced color, and the morphological correspondence with the broad 
molecular line
emitting regions, suggests that the infrared emission from clouds N and S is due 
to shocks in propagating into the clouds. Conversely, the normal infrared color 
of clouds A, B, and C is consistent with their being due to ambient molecular clouds.

If the infrared emission from clouds N and S is due to dust, then
the relatively higher $I_{60}/I_{100}$ could be due to smaller or warmer dust grains.
For dust heated by the average interstellar radiation field in the solar 
neighborhood, about half of the emission at 60 $\mu$m is thought to be due to
small, transiently heated grains (Draine \& Anderson 1985; D\'esert,
Boulanger, \& Puget 1990). Thus, if the enhanced 60 $\mu$m emission is
due to dust grains, then 
clouds N and S may contain a larger fraction of small
grains. An enhanced abundance of small grains would be expected if a significant
fraction of larger grains were shattered behind the shock front.
Observations of local cirrus clouds with
significant velocities revealed that $I_{60}/I_{100}\simeq 0.29$ is typical for clouds
with $V_{LSR}>30$ km~s$^{-1}$, suggesting that grain shattering was significant
in the shocks that accelerated local clouds to intermediate velocities (Heiles, 
Reach, \& Koo 1988).
Theoretically, significant shattering is not predicted for slow shocks such as 
inferred from the
widths of the CO lines, but faster shocks through somewhat lower-density 
interclump gas,
with $V_s>100$ km~s$^{-1}$, could produce the enhanced 60 $\mu$m emission
(Jones, Tielens, \& Hollenbach 1996).

Spectral lines could contribute significantly to the infrared emission from 
clouds N and S.
The most important lines in the {\it IRAS} passbands,
based on infrared spectra of similar supernova remnants
(Oliva et al. 1999, Cesarsky et al. 1999, Reach \& Rho 2000a),
are [\ion{O}{3}] 88~$\mu$m in {\it IRAS} band 4,
[\ion{O}{1}] 63~$\mu$m in band 3,
[\ion{Fe}{2}] 26 $\mu$m in band 2, and [\ion{Ne}{2}] 12.8 $\mu$m and 
H$_2$ lines in band 1.
If we were to interpret all of the {\it IRAS} emission
from cloud S as due to the ionic lines listed above, then,
using the system response and bandwidths (Beichman et al. 1988), we find that
the brightest line would be [\ion{O}{3}] 88 $\mu$m, 
with intensity $I=1.3\times 10^{-3}$ erg~s$^{-1}$~cm$^{-2}$~sr$^{-1}$.
Relative to this line, the other bright lines would
have ratios $\lambda\lambda$63/88=0.27, $\lambda\lambda$26/88=0.13, and
$\lambda\lambda$12.8/88=0.26.
The implied brightness of the [\ion{O}{1}] 63~$\mu$m line can be easily
produced by shocks with velocities $\sim 100$ km~s$^{-1}$ into
moderate-density ($10^3$ cm$^{-3}$) gas (Hollenbach \& McKee 1989); however, such shocks do 
not produce as much [\ion{O}{3}] emission as observed, because the column 
density
of highly-ionized gas is insufficient.
Slower shocks into denser gas can also produce [\ion{O}{1}] 63~$\mu$m lines
this bright (Draine, Roberge, \& Dalgarno 1983). 
However, slower (C-type) shocks produce essentially no ionic line 
emission---especially
not an ion such as \ion{O}{3}. Nor would the slow shocks destroy grains 
adequately to
produce significant gas-phase \ion{Fe}{2}. Therefore, if the infrared emission 
is
from slow shocks, the {\it IRAS} 100 $\mu$m band emission is from dust,
and the {\it IRAS} 60 $\mu$m band emission is from a mix of dust and the 
[\ion{O}{1}] 63 $\mu$m line. 
The nature of the {\it IRAS} 12 and 25 $\mu$m band 
emission is more difficult to constrain. If the infrared
emission is from slower shocks, such as inferred from the CO observations, then
there is likely a contribution from H$_2$ lines. 
H$_2$ rotational lines are the dominant coolant for a range of molecular
shocks (Rho et al. 2000, Reach \& Rho 2000a).
For now, it is not possible to clearly tell what fraction of the infrared 
emission
is from gas or dust. The nature of the infrared emission from HB~21 
(and other supernova remnants) can be determined in the future
using spectroscopy or narrow-band imaging.


\section{SUMMARY}

We have mapped the eastern half ($80'\times110'$) of the SNR HB~21 in $^{12}$CO J=2--1 line emission
almost completely.  Our map, 
which has been completely sampled with $27''$ resolution, shows the detailed structure of 
molecular clouds in this 
area. We have detected broad CO lines with large $\ratioto$
in the northern and southern parts of HB~21,
which is direct evidence for the interaction between molecular clouds and the SNR.
In the following, we summarize the main results of this paper:


\noindent
(1) We detected shocked molecular clouds, clouds N and S,  
with broad (20--40~\kms) CO lines 
in the northern and southern parts of the SNR.
Cloud N is composed of a small ($\sim 2'$ or 0.5~pc), very bright, 
U-shaped, clumpy part and several clumps scattered around it. 
Cloud S is filamentary and appears to form an elongated loop of $\sim 30'$ in extent.
Small ($\simlt 1'.2$ or 0.3~pc), bright clumps 
are seen along the filamentary structure. 
The H$_2$ masses of  
clouds N and S are $\sim 8$~\msol\ and $\sim 55$~\msol, respectively.\hb
(2) We have obtained sensitive J=1--0 and J=2--1 spectra of 
$^{12}$CO and $^{13}$CO molecules toward several peak positions of clouds 
N and S.
They have $\ratioto=1.6$--2.3 and $\ratiott=20$--40 
with CO J=2--1 main-beam brightness temperature less than 7~K.
According to our LVG analysis, $T_k\ge 50$~K, and, for $T_k=100$~K, 
$\nhtwod=$(3--7)$\times 10^3$~cm$^{-3}$ and 
$N({\rm CO})\simeq$(2.4--11) $\times 10^{17}$~cm$^{-2}$. The emitting region appears 
to fill a small (0.077--0.088) fraction of the beam.
\hb 
(3) There is an enhanced radio emission which attains a maximum 
exactly inside the 
central U-shaped part of cloud N. The emission has a spectral index 
($-0.28\pm0.17$) flatter than that 
of the whole remnant. The association of this 
emission with cloud N needs to be explored.\hb
(4) Clouds N and S are visible in the {\it IRAS} HIRES images at all four wavelengths 
(12, 25, 60, and 100~$\mu$m). 
They have the far-infrared color ratio $I_{60}/I_{100}\simeq 0.27$, which is significantly greater 
than that (0.20) of the other clouds in this area.  
This enhanced color, and the morphological correspondence with the broad 
molecular line
emitting regions, suggests that the infrared emission from clouds N and S is due 
to
shocks in propagating into the clouds.\hb
(5) Along the eastern boundary of the SNR, 
three relatively large ($\simgt 15'$ or 3.5~pc) clouds and 
several filamentary clouds are present. 
No broad CO emission or enhanced 60/100 $\mu$m color were 
detected in any of these clouds, and 
there is little relationship between the 
boundaries of the clouds and the SNR.
Therefore, there is no strong evidence for the interaction of the SNR with molecular clouds
along the eastern boundary.\hb

\acknowledgments
We thank Tom Landecker for kindly providing the radio images of HB 21.
B.-C. K. was supported by the S.N.U. Research Fund (99-9-2-041) and  
in part by the BK21 Project of the Korean Government.


{}

\clearpage

\figcaption[]
{$^{12}$CO J=2--1 integrated intensity map of HB~21. 
The velocity range is from 
$\vlsr=+3.9$~\kms\ to $-17.5$~\kms, 
and the integrated intensity varies from 0 to 64~K~\kms.
The overlaied contour map shows the 1420~MHz brightness
distribution of HB~21 obtained by T90 using the DRAO synthesis telescope. 
}

\figcaption[]
{A finding chart showing the positions of the sources described in the text.
The outermost box represents the area of our CO J=2--1 line observation while 
the inner square boxes represent the 
areas shown in Figures~3, 4, and 6. Thin solid contours show the boundary of CO distribution
while dashed contour shows the radio continuum boundary of the SNR HB~21.
}

\figcaption[]
{Channel maps of the eastern area. 
Central velocities are marked on the top of each map. 
Velocity widths are 3.3~\kms, 3.9~\kms, 4.6~\kms, 
and 5.2~\kms\ for maps a, b, c, and d, respectively.
Contour levels, which represent average brightness temperature in that 
velocity interval, are evenly spaced, with an 
interval of 1~K, starting at $\bar T_{mb}=1$~K.
Grey scale plot distinguishes hills from valleys.
}

\figcaption[]
{Same as Figure 3, but for the northern area. The crosses mark the peak positions, 
i.e., N1 and N2 from east to west, where $^{13}$CO lines have been obtained.
Velocity width of each map is 3.3~\kms. Contour levels are 
$\bar T_{mb}=$0.5, 1, 2, 3, and 4~K.
}

\figcaption[]
{Sensitive CO spectra of the northern peak positions N1 and N2.
The molecule and transition are marked in each
spectrum. The second spectrum from the top is
$^{12}$CO J=2--1 spectrum convolved to the J=1--0 beam size ($55''$) to be
compared with the J=1--0 spectra. 
The $^{12}$CO J=1--0 
spectrum in the right bottom frame, which is toward the position 
$4'$ away in RA from N1, is shown for comparison.
}

\figcaption[]
{Channel maps of the southern area. The crosses mark the peak positions, i.e.,
S1, S2, and S3 from east to west, where $^{13}$CO lines have been observed. 
Central velocities are marked on the top of each map. 
Velocity width of each map is 6.5~\kms.
Contour levels are 
$\bar T_{mb}=$0.75, 1.5, 3, 4.5, 6, and 7.5~K.
}

\figcaption[]
{Sensitive CO spectra of the southern peak positions S1, S2, and S3.
The molecule and transition are marked in each
spectrum. The second spectrum from the top is
$^{12}$CO J=2--1 spectrum convolved to the J=1--0 beam size ($55''$).
Dotted lines in $^{12}$CO J=2--1 spectra 
mark the velocity ($\vlsr=4$ to 5~\kms) where 
the spectra are corrupted by 
telluric CO emission.
}

\figcaption[]
{Result of our LVG model analysis for 
$T=100$~K. (a) Curves of constant $\ratioto$ (solid line) and 
constant $\ratiott$ (dotted line) are drawn. 
The asterisks (*) mark 
the observed ratios toward the peak positions where both
ratios are obtained, i.e., N1, N2, and S1.
(b) Curves of constant 
CO J=2--1 radiation temperature and
constant $\ratioot$ are drawn. The asterisks (*) mark ($\nhtwod$, $\xdvdr$) pairs determined 
from (a), i.e., they do {\it not} represent the observed 
radiation temperature and $\ratioot$.
}

\figcaption[]
{Enlarged view of the northern area. Grey-scale map shows the distribution of 
CO J=2--1 integrated intensity while contour map shows the distribution of 
325 MHz continuum brightness obtained from the Westerbork Northern 
Sky Survey (Rengelink et al. 1997). 
Note that the position of radio continuum peak coincides 
with broad-line emitting CO cloud.
Contour levels are evenly spaced, with an 
interval of 0.01~K, starting at $\bar T_{mb}=0.01$~K.
}

\figcaption[]
{The 60~$\mu$m image of HB~21. The CO distribution is shown in solid contours.
}

\clearpage

\begin{deluxetable}{ccccccc}
\scriptsize
\tablecaption{Observed Parameters at the Peak Positions with Broad Emission Lines
\label{tbl-1}}
\tablewidth{0pt}
\tablecolumns{7}
\tablehead{\colhead {Name} 
& \colhead { ({$\alpha_{\rm 1950}$}, {$\delta_{\rm 1950}$}) }
& \colhead {($v_{\rm min}$, $v_{\rm max}$)}
& \colhead {$T_{\rm max}$}
& \colhead {$\ratioto$} 
& \colhead {$\ratiott$} 
& \colhead {$\ratioot$} \\
\colhead {} & \colhead {(\ h\ \  m\ \  s, \ \ $^\circ$\ \ \ $'$\ \ \ $''$)} 
& \colhead {(\kms)} & \colhead {(K)} & \colhead {} & \colhead {} &\colhead {} }
 
\startdata
 
HB21:BML-N1 & (20 46 03.2, 51 00 00) & $(-20, +10)$ &	3.6 &  1.6&$20\pm 3$&$20\pm 1$\nl
HB21:BML-N2 & (20 45 55.0, 51 03 30) & $(-21, +11)$ &	3.1 &  1.7&$40\pm14$&...\nl
HB21:BML-S1 & (20 44 37.2, 49 47 10) & $(-22, +20)$ &	2.7 &  1.9&$28\pm 4$&$96\pm 50$\nl
HB21:BML-S2 & (20 44 31.0, 49 55 20) & $(-27,  +0)$ &	2.9 &  1.7&...&$104\pm 43$\nl
HB21:BML-S3 & (20 42 45.2, 49 56 50) & $(-35, +1)$  &	6.9 &  2.3&...&$103\pm 43$\nl
 
\enddata
 

\end{deluxetable}

\clearpage 
 
\begin{deluxetable}{cccccc}
\tablecaption{Physical Parameters Derived from LVG Analysis
\label{tbl-2}}
\tablewidth{0pt}
\tablecolumns{6}
\tablehead{ 
\colhead {Name}
& \colhead {$\nhtwod$} 
& \colhead {$\xco/(dv/dr)$} 
& \colhead {$N({\rm CO})$} 
& \colhead {beam-filling}
& \colhead {$\ratioot$} \\
\colhead {} 
& \colhead {($10^3$~cm$^{-3}$)} 
& \colhead {(10$^{-6}$ pc/km$^{-1}$ s)}
& \colhead {($10^{17}$~cm$^{-2}$)}
& \colhead {factor} & \colhead {} }
 
\startdata
 
N1 & 6.1 & 4.3 & 11 & 0.077 & 40 \nl
N2 & 3.1 & 1.4 & 2.4 & 0.088 &57 \nl
S1 & 7.0 & 2.1 & 6.4 & 0.079 & 50 \nl
 
\enddata
\end{deluxetable}
\clearpage 

\begin{deluxetable}{cccccc}
\tablecaption{Infrared Properties of Molecular Clouds
\label{tbl-3}}
\tablewidth{0pt}
\tablecolumns{6}
\tablehead{ 
\colhead {Cloud}
& \colhead {$I_{12}$} 
& \colhead {$I_{25}$} 
& \colhead {$I_{60}$} 
& \colhead {$I_{100}$} 
& \colhead {$I_{60}/I_{100}$} \\
\colhead {Name} 
& \colhead {(MJy sr$^{-1}$)} 
& \colhead {(MJy sr$^{-1}$)} 
& \colhead {(MJy sr$^{-1}$)} 
& \colhead {(MJy sr$^{-1}$)} 
& \colhead {} }
 
\startdata
 
A &      1.1(0.2) & 1.1(0.1) & 4.2(0.2) & 20.8(2.3) & 0.20(0.03)\nl
B &      1.7(0.1) & 1.8(0.1) & 5.9(1.1) & 29.5(2.4) & 0.20(0.04)\nl
C &      1.7(0.1) & 1.7(0.1) & 8.0(1.1) & 38.0(2.4) & 0.21(0.03)\nl
N &      1.0(0.2) & 0.8(0.1) & 4.8(0.2) & 17.0(2.3) & 0.28(0.04)\nl
S &      1.2(0.1) & 1.2(0.1) & 6.8(1.1) & 26.3(2.4) & 0.26(0.05)\nl

\enddata
\tablenotetext{}{NOTE.---Numbers in parentheses are estimated (1$\sigma$) errors.}

\end{deluxetable}


\clearpage

\begin{thebibliography}{}

%
\bibitem[]{} Arendt, R. G. 1989, ApJS, 70, 181
\bibitem[]{} Arikawa, Y., Tatematsu, K., Sekimoto, Y., \& Takahashi, T. 1999, PASJ, 51, L7
\bibitem[Aumann, Fowler, \& Melnyk 1990]{hiresref} Aumann, H. H., 
Fowler, J. W., \& Melnyk, M. 1990, AJ, 99, 1674
\bibitem[Beichman et al. 1988]{irassup} Beichman, C. A. et al. 1988, {\it Infrared
Astronomical Satellite (IRAS) Catalogs and Atlases: Volume 1. Explanatory
Supplement}, NASA RP-1190 (NASA: Washington, DC)
\bibitem[]{} Blandford, R. D., \& Cowie, L. L. 1982, ApJ, 260, 625
\bibitem[]{} Boulanger, F., \& P\'erault, M. 1988, ApJ, 330, 964
\bibitem[]{} Cesarsky, D., Cox, P., Pineau Des For\^ets, G., van Dishoeck, E. F., Boulanger, 
F., \& Wright, C. M. 1999, A\& A, 348, 945
\bibitem[]{} Chevalier, R. A. 1974, ApJ, 188, 501
\bibitem[]{} ------ 1999, ApJ, 511, 798
\bibitem[]{} DeNoyer, L. K. 1979, ApJL, 232, L165
\bibitem[]{} D\'esert, F. X., Boulanger, F., \& Puget, J.-L. 1990, A\&A, 237, 215
\bibitem[]{} Dickman, R. L., Snell, R. L., Ziurys, L. M., \& Huang, Y.-L. 1992,
ApJ, 400, 203
\bibitem[]{} Dobashi, K., Bernard, J.-P., Yonekura, Y., \& Fukui, Y. 1994, ApJS, 95, 419
\bibitem[]{} Draine, B. T., \& Anderson, N. 1985, ApJ, 292, 494
\bibitem[]{} Draine, B. T., \& McKee, C. F. 1993, ARAA, 31, 373
\bibitem[]{} Draine, B. T., Roberge, W. G., \& Dalgarno, A. 1983, ApJ, 264, 485
\bibitem[]{} Draine, B. T., \& Roberge, W. G. 1984, ApJ, 282, 491
\bibitem[]{} Erkes, J. W., \& Dickel, J. R. 1969, AJ, 74, 840
\bibitem[]{} Frail, D. A., Goss, W. M., Reynoso, E. M., Giacani, E. B., 
Green, A. J., \& Otrupcek, R. 1996, AJ, 111, 1651
\bibitem[]{} Goldreich, P., \& Kwan, J. 1974, ApJ, 189, 441
\bibitem[]{} Green, D. A. 1986, MNRAS, 221, 473
\bibitem[]{} Heiles, C., Reach, W. T., \& Koo, B.-C. 1988, ApJ, 332, 313
\bibitem[]{} Hill, I. E. 1974, MNRAS, 169, 59
\bibitem[]{} Hollenbach, D., \& McKee, C. F. 1980, ApJL, 241, L47
\bibitem[Hollenbach \& McKee 1989]{hm89} ------
1989, ApJ, 342, 306
\bibitem[]{} Huang, Y.-L., \& Thaddeus, P. 1986, ApJ, 309, 804
\bibitem[]{} Humphreys, R. M. 1978, ApJS, 38, 309
\bibitem[]{} Jones, T. W., \& Kang, H. 1993, ApJ, 402, 560
\bibitem[]{} Jones, A. P., Tielens, A. G. G. M., \& Hollenbach, D. J. 1996, ApJ, 469, 740
\bibitem[]{} Keohane, J. W., Petre, R., Gotthelf, E. V., Ozaki, M., \& Koyama, K. 1997, ApJ, 484, 350
\bibitem[]{} Klein, R. L., McKee, C. F., Colella, P. 1994, ApJ, 420, 213
\bibitem[]{} Koo, B.-C., \& Heiles, C. 1991, ApJ, 382, 204
\bibitem[]{} Koo, B.-C. \& Moon, D.-S. 1997, ApJ, 485, 263
\bibitem[]{} Langer, W. D. 1997, in CO: Twenty-Five Years of Millimeter-Wave Spectroscopy,
IAU Symp. 170, ed. Latter, W. B. et al. (Dordrecht: Kluwer), 98
\bibitem[]{} Leahy, D. A. 1987, MNRAS, 228, 907
\bibitem[]{} Leahy, D. A., \& Aschenbach, B. 1996, A\&A, 315, 260
\bibitem[]{} Lockett, P., Gautheir, E., Elitzur, M. 1999, ApJ, 511, 235
\bibitem[]{} Mac Low, M.-M., McKee, C. F., Klein, R. I., Stone, J. M., \& Norman, M. L. 
1994, ApJ, 433, 757
\bibitem[]{} McKee, C. F. 1988, in Supernova Remnants and the Interstellar
Medium, Proceedings of IAU Colloquium No. 101, ed. R.S. Roger and T.L.
Landecker (Cambridge: Cambridge), 473
\bibitem[]{} Oliva, E., Moorwood, A. F. M., Drapatz, S., Lutz, D., \& Sturm, E. 1999, A\& A, 
343, 943
\bibitem[]{} Reach, W. T., \& Rho, J. 1996, A\&A, 315, L277
\bibitem[]{}------. 1999, ApJ, 511, 836
\bibitem[Reach \& Rho 2000b]{rr00} ------. 2000a, Aph0007148
\bibitem[Reach \& Rho 2000a]{} ------. 2000b, in preparation
\bibitem[]{} Rengelink, R. B., Tang, Y., de Bruyn, A. G., Miley, G. K., Bremer, M. N., R\"ottgering, H. J. A., \& Bremer, M. A. R. 1997, A\&AS, 124, 259
\bibitem[]{} Rho, J., \& Petre, R. 1998, ApJL, 503, L167
\bibitem[Rho et al. 2000]{rho443} Rho, J., Jarrett, T., Cutri, R. M., 
\& Reach, W. T. 2000, ApJ, submitted
\bibitem[]{} Roh, D.-G., \& Jung, J. H. 1999, Publications of the Korean Astronomical 
Society, 14, 123
\bibitem[]{} Sakamoto, S., Hasegawa, T., Handa, T., Hayashi, M., \& Oka, T. 1997, ApJ, 486, 276
\bibitem[]{} Sakamoto, S., Hayashi, M., Hasegawa, T., Handa, T., \& Oka, T. 1994, ApJ, 425, 641
\bibitem[]{} Scoville, N. Z., \& Solomon, P. M. 1974, ApJL, 187, L67
\bibitem[]{} Seta, M., Hasegawa, T., Dame, T. M., Sakamoto, S., 
Oka, T., Handa, T., Hayashi, M., Morino, J., Sorai, K., \& Usuda, K. S.
 1998, ApJ, 505, 286 
\bibitem[]{} Tatematsu, K., Fukui, Y., Landecker, T. L., \& Roger, R. S. 1990, A\&A, 237, 189 (T90)
\bibitem[]{} Tauber, J. A., Snell, R. A., Dickman, R. L., \& Ziurys, L. M. 1994,
      ApJ, 421, 570
\bibitem[]{} van den Bergh, S. 1978, ApJS, 38, 119
\bibitem[]{} van der Laan, H. 1962, MNRAS, 124, 179
\bibitem[]{} van Dishoeck, E. F., Jansen, D. J., \& Phillips, T. G. 1993, A\&A, 279, 541
\bibitem[Wheelock et al. 1993]{issaref} Wheelock S. L., Gautier, T. N., Chillemi, J., Kester,
D., McCallon, H., Oken, C., White, J., Gregorich, D., Boulanger, F.,
and J. Good 1994, {\it IRAS Sky Survey Atlas: Explanatory Supplement}.
(JPL/Caltech: Pasadena)
\bibitem[]{} Willis, A. G., 1973, A\&A, 26, 237
\bibitem[]{} Wilner, D. J., Reynolds, S. P., \& Moffett, D. A. 1998, AJ, 115, 247


\end{thebibliography}
\end{document}